# Using Evolution Strategy with Meta–models for Well Placement Optimization

Zyed Bouzarkouna (IFP), Didier Yu Ding (IFP) & Anne Auger (INRIA)

**Abstract**

Optimum implementation of non-conventional wells allows us to increase considerably hydrocarbon recovery. By considering the high drilling cost and the potential improvement in well productivity, well placement decision is an important issue in field development. Considering complex reservoir geology and high reservoir heterogeneities, stochastic optimization methods are the most suitable approaches for optimum well placement. This paper proposes an optimization methodology to determine optimal well location and trajectory based upon the Covariance Matrix Adaptation - Evolution Strategy (CMA-ES) which is a variant of Evolution Strategies recognized as one of the most powerful derivative-free optimizers for continuous optimization. To improve the optimization procedure, two new techniques are investigated: (1). Adaptive penalization with rejection is developed to handle well placement constraints. (2). A meta-model, based on locally weighted regression, is incorporated into CMA-ES using an approximate ranking procedure. Therefore, we can reduce the number of reservoir simulations, which are computationally expensive. Several examples are presented. Our new approach is compared with a Genetic Algorithm incorporating the Genocop III technique. It is shown that our approach outperforms the genetic algorithm: it leads in general to both a higher NPV and a significant reduction of the number of reservoir simulations.

## 1. Introduction

A well placement decision affects significantly the hydrocarbon recovery and thus the asset value of the project. In general, such a decision is difficult to make since optimal placements depend on a large number of parameters such as reservoir heterogeneities, formation characteristics... Moreover, dealing with complex well configurations, i.e., non-conventional wells, implies several challenges: the concentration of investment, the well intervention difficulty...

Recently, many efforts have been made to formulate the well placement decision as an optimization problem which aims to offer an assisted tool insuring a higher probability to find the optimal well configuration with respect to a given objective function describing the economics of the project, e.g., the Net Present Value (NPV).

In the publications related to well placement optimization, two main optimization approaches have been mostly used: gradient based methods and genetic algorithms. The use of gradient based methods (e.g., Handels et al., 2007 and Sarma & Chen, 2008) was motivated by the fact that it is by far less reservoir simulations consuming compared to stochastic methods. However, these deterministic methods find a difficulty in general when dealing with multi-modal, non-smooth and non-convex problems such as the well placement problem.

On the other hand, genetic algorithms, a specific type of Evolutionary Algorithm (EA), have received the most important part of interest. Genetic algorithms (e.g., Montes et al., 2001 and Artus et al., 2006), as well as the majority of stochastic algorithms, have the ability to deal with difficult functions to solve. Moreover, the problem of the computational effort can be tackled by combining stochastic algorithms with acceleration routines (Guyaguler & Horne, 2000, Guyaguler et al., 2000, Guyaguler & Horne, 2001, Yeten et al., 2003 and Emerick et al., 2009).

This paper presents a new assisted methodology for well placement optimization based on a real representation stochastic method: the Covariance Matrix Adaptation – Evolution Strategy (CMA-ES). The choice of CMA-ES was motivated by the fact that CMA-ES is recognized as one the most powerful methods for derivative-free optimization (Hansen et al., 2010). A first application of CMA-ES to well placement optimization was introduced by Ding (2008).
To improve the optimization procedure for well placement with CMA-ES, two new techniques are introduced:

- A method to handle constraints using adaptive penalization with rejection;
- A meta-model approach based on a locally weighted regression and incorporated into CMA-ES using an approximate ranking procedure, to reduce the number of reservoir simulations, which are very demanding in terms of CPU time.

This paper is structured as follows. In Section 2, we introduce the optimization algorithm CMA-ES. Handling constraints is tackled using adaptive penalization with rejection. In Section 3, CMA-ES is compared to a genetic algorithm on a synthetic reservoir case to show the contribution of this optimization method. In Section 4, the reduction of the number of reservoir simulations is also addressed by coupling CMA-ES with meta-models. Finally, in Section 5, the contribution of the proposed methodology is demonstrated on a well location and trajectory optimization problem.

## 2. Covariance Matrix Adaptation ES

*Covariance Matrix Adaptation-ES* CMA-ES (Hansen & Ostermeier, 2001) is a stochastic optimization algorithm where at each iteration $g$, a population of $\lambda$ points is sampled according to a multivariate normal distribution. The objective function, denoted by $f$, of the $\lambda$ points is then evaluated and the parameters of the multivariate normal distribution are updated using the feedback obtained on the objective function. More specifically, let **m** be the mean value of the multivariate normal distribution generated by CMA-ES, constituting an estimate of the optimum and let $\sigma$ and **C** be respectively the step-size and covariance matrix. Assume that **m**, σ, **C** are given, new points or *individuals* are sampled according to:

$$\mathbf{x}_i = \mathbf{m} + \sigma \aleph_i(0, \mathbf{C}) \ , \forall \ i = 1,...,\lambda , \tag{1}$$

where $(\aleph_i(0, \mathbf{C}))_{1 \leq i \leq \lambda}$ are λ independent multivariate normal distributions with zero mean vector and covariance matrix **C**. Those λ individuals are ranked according to *f*:

$$f(\mathbf{x}_{1:\lambda}) \leq ... \leq f(\mathbf{x}_{\mu:\lambda}) \leq ... \leq f(\mathbf{x}_{\lambda:\lambda}) \tag{2}$$

where we use the notation $\mathbf{x}_{i:\lambda}$ for the $i^{th}$ best individual.

The mean **m** is then updated by taking the weighted mean of the best μ individuals:

$$m \leftarrow \sum_{i=1}^{\mu} \omega_i \mathbf{x}_{i:\lambda} \ , \tag{3}$$

where in general $\mu = \dfrac{\lambda}{2}$ and $(\omega_i)_{1 \leq i \leq \lambda}$ are strictly positive and normalized weights, i.e., satisfying $\sum_{i=1}^{\mu} \omega_i = 1$.

Furthermore the covariance matrix of the search distribution **C** and the step-size σ are updated as well for the next generation and we refer to Hansen & Kern (2004) for the equation updates.
All updates rely only on the ranking determined by Eq. 2 and not on the exact value of the objective functions.

*Handling Constraints with CMA-ES* Several methods are used, in the literature, to handle constraints in evolutionary algorithms. Unfeasible individuals can be rejected, penalized or repaired... In the following, we briefly discuss these alternatives. A more detailed study and comparison can be found in Michalewicz et al. (1996).

- Rejection of unfeasible individuals: Besides its simplicity and ease of implementation, applying the "death penalty" does not require any parameter to be tuned. However, ignoring unfeasible individuals can prevent the algorithm from finding the region containing the global optimum if this one is close to the feasible domain boundaries;

- Penalizing unfeasible individuals: Penalization is the most widespread approach used to handle constraints. This method corresponds to a transformation of the optimization problem:

$$\begin{cases} \min f(\mathbf{x}) \\ \text{s.t.} \ h_i(\mathbf{x}) \leq d_i \ \forall i = 1,...,m \end{cases} \Rightarrow \min f(\mathbf{x}) + \sum_{i=1}^{m} g(h_i(\mathbf{x}) - d_i) \ , \tag{4}$$

where *m* is the number of constraints and *g*(.) is the penalty function which is non-negative equal to zero in IR_ and increasing in IR_+. In general, *g*(.) contains parameters to be tuned. These parameters depend on the problem to be optimized. A solution to avoid this problem consists on using an adaptive penalization which does not require any user specified constants. However, penalizing all unfeasible individuals implies evaluating all individuals which is in general costly in terms of CPU time, due to the cost of the objective function evaluation;

- Repairing unfeasible individuals: Another popular solution to handle constraints is to repair each unfeasible individual before evaluating it. An important parameter to be specified is the probability of the replacement of the unfeasible individual by the repaired new feasible individual. Moreover, every time the algorithm finds an unfeasible individual, repair introduces a repaired individual that does not obey necessarily to the already adapted covariance matrix, which may hold up the optimization process of CMA-ES.

Knowing the limitations of each of the constraint-handling approaches, the approach used in the present paper is a mixture between 2 approaches: *adaptive penalizing marginally unfeasible individuals* and *rejecting only unfeasible individuals far from the boundaries of the feasible domain*. Using this approach, rejecting only individuals far from the feasible domain does not prevent the algorithm from finding a solution near the constraint boundary, and by using adaptive penalization, the algorithm process becomes problem-independent.

Hansen et al. (2009) present a box constraint handling, in which the feasible space is a hypercube defined by lower and upper boundary values for each parameter. In the present paper, this approach is generalized in order to handle feasible spaces defined by a lower and upper boundary values for a sum of some of the parameters.

Given an optimization problem with a dimension $n$, let us suppose we have $m \in \mathbf{N}$ constraints denoted by $S_j$, $\forall j = 1,...,m$. For each constraint $S_j$, we define $P_j \subset \{1..n\}$ such that a vector $\mathbf{x} = (x_i)_{1 \leq i \leq n}$ is feasible with respect to the constraint $S_j$ if:

$$v_{(j,-)} < q_j = \sum_{p \in P_j} x_p < v_{(j,+)} , \tag{5}$$

where $v_{(j,-)}$ and $v_{(j,+)}$ are respectively the lower and upper boundaries defining the constraint $S_j$.

Penalizing is performed as follows, when evaluating an individual $\mathbf{x}$:

- *Initializing weights*: In the first generation, boundary weights $\gamma_j$ are initialized. $\gamma_j = 0$, $\forall j = 1,...,m$;

- *Setting weights*: From the second generation upwards, if the distribution mean is unfeasible and weights are not set yet, then

$$\gamma_j \leftarrow \frac{2\delta_{\text{fit}}}{\sigma^2 \frac{1}{n}\sum_{i=1}^{n} C_{ii}}, \forall j = 1,...,m , \tag{6}$$

where $\delta_{\text{fit}}$ is the median from the last $(20 + 3n)/\lambda$ generations of the interquartile range of the unpenalized objective function evaluations and $C_{ii}$ is the $i^{\text{th}}$ diagonal element of the covariance matrix $\mathbf{C}$ generated by CMA-ES;

- *Increasing weights*: For each constraint $S_j$, if the distribution mean $m_j$, i.e., the mean of $q_j$ for the $\lambda$ individuals of the current generation, is out-of-bounds and the distance from $m_j$ to the feasible domain, i.e., $\max(0, m_j - v_{(j,+)}) + \max(0, v_{(j,-)} - m_j)$ is larger than $\sigma \times \sqrt{\frac{1}{card(P_j)} \sum_{p \in P_j} C_{pp}} \times \max(1, \frac{\sqrt{n}}{\mu_{eff}})$

$$\gamma_j \leftarrow \gamma_j \times 1.1^{\max(1, \frac{\mu_{\text{eff}}}{10n})}, \forall j = 1,...,m , \tag{7}$$

- *Evaluating the individual*:

$$f(\mathbf{x}) \leftarrow f(\mathbf{x}) + \frac{1}{m} \sum_{j=1}^{m} \gamma_j \frac{(q_j^{\text{feas}} - q_j)^2}{\xi_j} , \tag{8}$$

where $q_j^{\text{feas}}$ is the projection of $q_j$ on the feasible domain with respect to the constraint $S_j$, $card(P_j)$ is the number of elements in the set $P_j$ and $\xi_j = \exp\left(0.9\left(\frac{1}{card(P_j)} \sum_{p \in P_j} \log(C_{pp}) - \frac{1}{n}\sum_{i=1}^{n} \log(C_{ii})\right)\right)$.

An individual $\mathbf{x}$, in the following, will be rejected and resampled if $\|q_j^{\text{feas}} - q_j\| > p \times \|q_j\|$, where $p$ is a parameter to be chosen. In all considered cases in this paper, $p$ was chosen to be equal to 20%.

## 3. Well placement optimization using CMA-ES

***Why CMA-ES?*** The choice of a stochastic optimization method was motivated by the ability of this type of algorithms to deal with non-smooth, non-convex and multi-modal functions. In addition, stochastic optimization does not require any gradients and can be easily parallelized. CMA-ES has been recognized as one of the most powerful continuous optimization algorithms on benchmark problems (Hansen et al., 2010) and real-world problems. However, genetic algorithms, especially binary encoded genetic algorithms, are still the most popular field in evolutionary algorithms used in the well placement optimization problem.

To motivate our choice, a comparison between CMA-ES and the genetic algorithm was performed on a well placement problem. The considered genetic algorithm uses a real representation, since binary representations genetic algorithms have many drawbacks when dealing with continuous optimization problems (Surry & Radcliffe, 1996).

*Well parameterization* The optimization problem is formulated so as to handle different possible types of multilateral wells. The mainbore is defined by a sequence of points $(\mathbf{P}_{d,i})_{0 \leq i \leq n\text{ deviations}}$. The starting point $\mathbf{P}_0$ of the mainbore called the heel is represented by its Cartesian coordinates $(x_0, y_0, z_0)$. Other intermediate points $(\mathbf{P}_{d,i})_{1 \leq i \leq n\text{ deviations-1}}$ and the ending point $\mathbf{P}_{d,i=n\text{ deviations}}$ called the toe are represented by their corresponding spherical coordinate system $(r_{d,i}, \theta_{d,i}, \varphi_{d,i})$ with respect to the basis $(\mathbf{P}_{d,i-1}, \mathbf{u}_{d,i}^r, \mathbf{u}_{d,i}^\theta, \mathbf{u}_{d,i}^\varphi)$.

The formulation of the optimization problem handles also a number $n_{\text{branches}}$ of branches and/or laterals. The branch or lateral $j \in [1,..., n_{\text{branches}}]$ is defined by locating its ending point $\mathbf{P}_{b,j}$ ($l_{b,j}$, $r_{b,j}$, $\theta_{b,j}$, $\varphi_{b,j}$) where $(r_{b,j}, \theta_{b,j}, \varphi_{b,j})$ represents the spherical coordinates of $\mathbf{P}_{b,j}$ with respect to the basis $(\mathbf{P}_{\cap,j}, \mathbf{u}_{b,j}^r, \mathbf{u}_{b,j}^\theta, \mathbf{u}_{b,j}^\varphi)$, $\mathbf{P}_{\cap,j}$ is the starting point of the branch or the lateral $j$, and $l_{b,j}$ is the length between $\mathbf{P}_0$ and $\mathbf{P}_{\cap,j}$.

The dimension $\dim_w$ of the representation of a well $w$ is as follows:
$$\dim_w = 3 \times (1 + n_{\text{deviations}}^w) + 4 \times n_{\text{branches}}^w, \qquad (9)$$

Hence, the dimension dim of the problem of placing $n_{\text{wells}}$ wells $(w_k)_{k=1...n\text{wells}}$ is:
$$\dim = \sum_{k=1}^{nwells} \dim_{wk}, \qquad (10)$$

In this paper, we suppose that the number of injectors and producers to be placed is an already defined data of the problem.

*Well placement using CMA-ES* The population size $\lambda$ is an important parameter of CMA-ES. It is true that choosing a large $\lambda$ may prevent the optimization process from being stuck in a local optimum but it may also lead to an important increase in the number of reservoir simulations performed. In this paper, tuning optimally the population size was not addressed. However, a number of runs were performed to choose a suitable population size for the problem at hand. The initial population is normally drawn using a mean vector uniformly drawn in the reservoir. A maximum number of iterations is fixed to 100. Other parameters were defined as standard settings in Hansen & Kern (2004). The objective function is defined by the NPV as in Yeten (2003):

$$\text{NPV} = \sum_{n=0}^{Y} \left( \frac{1}{(1+\text{APR})^n} \begin{bmatrix} Q_{n,o} \\ Q_{n,g} \\ Q_{n,w} \end{bmatrix}^T \begin{bmatrix} C_{n,o} \\ C_{n,g} \\ C_{n,w} \end{bmatrix} \right) - C_d, \qquad (11)$$

where $Q_{n,p}$ is the field production of phase $p$ in the period $n$, $C_{n,p}$ is the profit or loss associated with this production. A phase $p$ is either oil, gas or water denoted respectively $o$, $g$, $w$. APR is the annual discount rate. $Y$ is the number of discount period. $C_d$ is the drilling and completing cost for all wells approximated by:

$$C_d = \sum_{k=0}^{N_{lat}} [\text{A}.d_w.\ln(l_w).l_w]_k + \sum_{m=1}^{N_{jun}} [C_{jun}]_m, \qquad (12)$$

Eq. 12 is defined based on the approximate formula used by Yeten (2003). In eq. 12, $k = 0$ represents the mainbore, $k > 0$ represents the laterals, $l_w$ is the length of the lateral (in ft), $d_w$ is the diameter of the mainbore (in ft) and A is a constant specific to the considered field. $C_{jun}$ is the cost of milling the junction. Different constants in eq. 11 and eq. 12 were defined in Tab. 1.

**Tab. 1**: Constants used to define the Net Present Value (NPV).

| Constant | Value |
|---|---|
| $C_{n,o}$ | 60 \$ / bbl |
| $C_{n,w}$ | -4 \$ / bbl |
| $C_{n,g}$ | 0 |
| APR | 0 |
| A | 1000 |
| $d_w$ | 0.1 m |
| $C_{jun}$ | $10^5$ \$ |

**Tab. 2**: GA parameters: the application probabilities of the used operators, i.e., crossover and mutation.

| Constant | Value |
|---|---|
| *crossprob* | 0.7 |
| *mutprob* | 0.1 |

Different constraints were implemented to define the feasible domain of the optimization. The constraints handled in the present paper are as follows:
- maximum length of wells: $l_w < L_{max}$, $\forall w = 1,...,n_{wells}$;
- all wells must be inside the reservoir grid: we suppose that a well is inside the reservoir if its all corresponding points are inside the reservoir.

***Well placement using a Genetic Algorithm (GA)*** In order to motivate our choice of using CMA-ES, a comparison between CMA-ES and a genetic algorithm is performed.

Genetic algorithms are stochastic search algorithms that borrow some concepts from nature. Similar to CMA-ES, GAs are based on an initial population of individuals. Each individual represents a possible solution to the problem at hand. Starting with an initial population of points called individuals or *chromosomes*, and at each iteration, candidate solutions evolve by *selection*, *mutation* and *recombination* until reaching the stopping criteria with a satisfactory solution. The correspondence between a solution and its representation is defined. Usually simple forms like an array or a matrix of integer or bit elements are used. In this paper, individuals are parameterized as defined for CMA-ES. Hence, well coordinates are defined using a real encoding. Elitism is used to make sure that the best chromosome would survive to the next generation. The used operators are defined as follows:

- The crossover: The crossover starts with two parent chromosomes causing them to unite in points to create two new elements. The greater chromosome fitness's rank, the higher probability it will be selected. After selecting the two parents, crossover is applied with a crossover probability denoted *crossprob*. To apply the crossover, we randomly draw an index *i* between 1 and *dim* and a real *c* between 0 and 1. Let us denote the first parent by $(x_{1,j})_{1 \leq j \leq dim}$ and the second parent by $(x_{2,j})_{1 \leq j \leq dim}$.

    By performing crossover $x_{1,i} \leftarrow c \times x_{1,i} + (1-c) \times x_{2,i}$ and $x_{2,i} \leftarrow c \times x_{2,i} + (1-c) \times x_{1,i}$.

- The mutation: The mutation, instead, starts with one individual and randomly changes some of its components. Mutation is applied to all chromosomes, except the one with the best fitness value, with a probability of mutation denoted *mutprob*. In this case, we randomly draw an index *i*. Let us denote the selected chromosome by $(x_j)_{1 \leq j \leq dim}$.

    By performing mutation $x_i \leftarrow \min_i + c \times (\max_i - \min_i)$, where $\min_i$ and $\max_i$ are the minimum and the maximum values that can be taken by the $i^{th}$ coordinate of the chromosome.

The application probabilities for these operators are typical parameters for a genetic algorithm and were defined in Tab. 2.

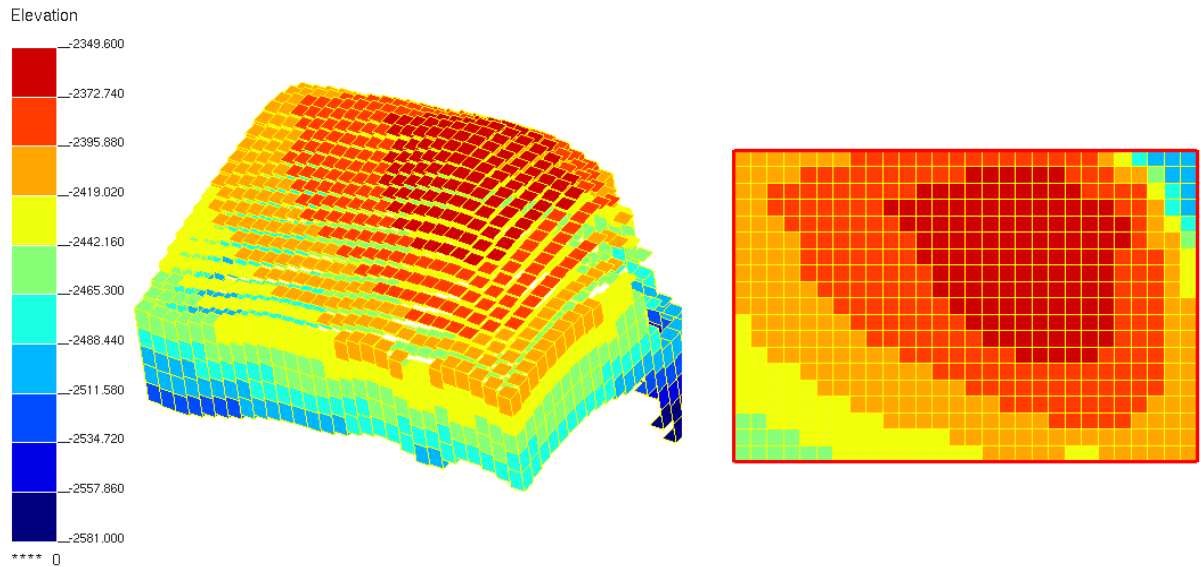

**Fig. 1**: PUNQ-S3 test case.

As the number of generations increases, fitter and fitter solutions are likely to be found. This process continues until satisfying the stopping criterion.

To handle the same constraints as with CMA-ES, the genetic algorithm is combined with the Genocop III technique (Genetic Algorithm for Numerical Optimization of Constrained Problems) as used by Emerick et al. (2009). Genocop III maintains 2 separate populations. The first population called a search population contains individuals satisfying linear constraints. The second population, a reference population, consists of individuals satisfying all constraints (linear and non-linear). Feasible search and reference individuals are evaluated directly using the objective function. However, unfeasible individuals are repaired before being evaluated.
More details about Genocop III can be found in (Michalewicz & Nazhiyath, 1995).

*Test case: PUNQ-S3* Tests were performed on the PUNQ-S3 test case. The PUNQ-S3 case has been taken from a reservoir engineering study on a real field. The model contains 19×28×5 grid blocks. We suppose that the field does not contain any production or injection wells initially. The elevation of the field is shown in Fig. 1. We propose to place 2 wells: 1 unilateral water injector and 1 unilateral producer to maximize the NPV. The dimension of the problem is then equal to: $6 \times 2 = 12$.

*Well placement performance* To compare results obtained by both CMA-ES and the genetic algorithm, 14 runs were performed for each method. A streamline simulator is used during the optimization. In this comparison, a producer limit bottomhole pressure is fixed to 80 bar, and an injector limit bottomhole pressure is fixed to 6.000 bar which is too high. These impractical values were chosen only for the sake of comparison between the two optimization methods.
The population size for both algorithms is set to 40. The size of the reference population for Genocop III is chosen to be equal to 60.

Fig. 2 shows the average performance of well placement optimization using both methods measured by the overall best objective function value. It is clear that CMA-ES outperforms the GA: the genetic algorithm improves only by 40% to the best NPV obtained by a random configuration, i.e., in the first generation of the optimization. However, CMA-ES improves it by 80%.

Fig. 3 shows that CMA-ES handles the considered constraints successfully. The number of well configurations resampled, i.e., far from the feasible domain, approaches 0 at the end of the optimization. Fig. 3 shows that after a number of iterations, all well configurations generated by CMA-ES are either feasible or close to the feasible domain.

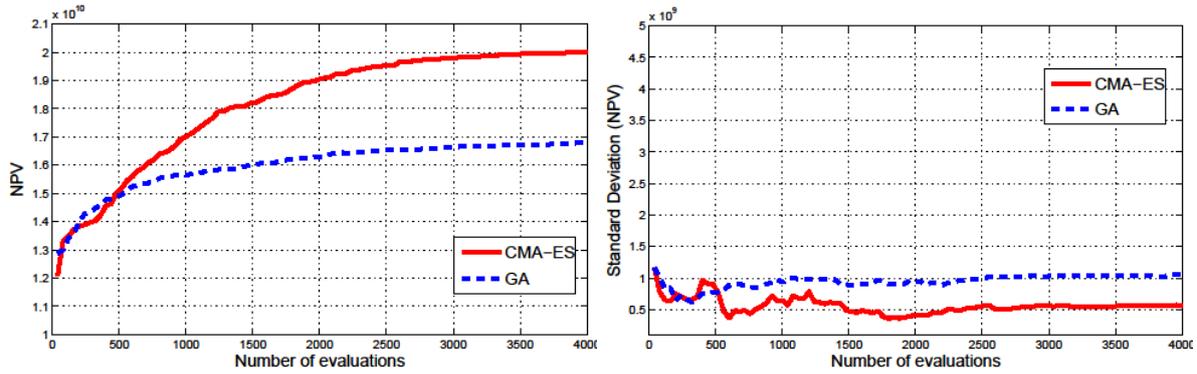

**Fig. 2**: The mean value of NPV (in US$) and its corresponding standard deviation for well placement optimization using CMA-ES (———) and GA (- -). 14 runs are performed for each algorithm. Constraints are handled using an adaptive penalization with rejection technique for CMA-ES and using Genocop III for GA.

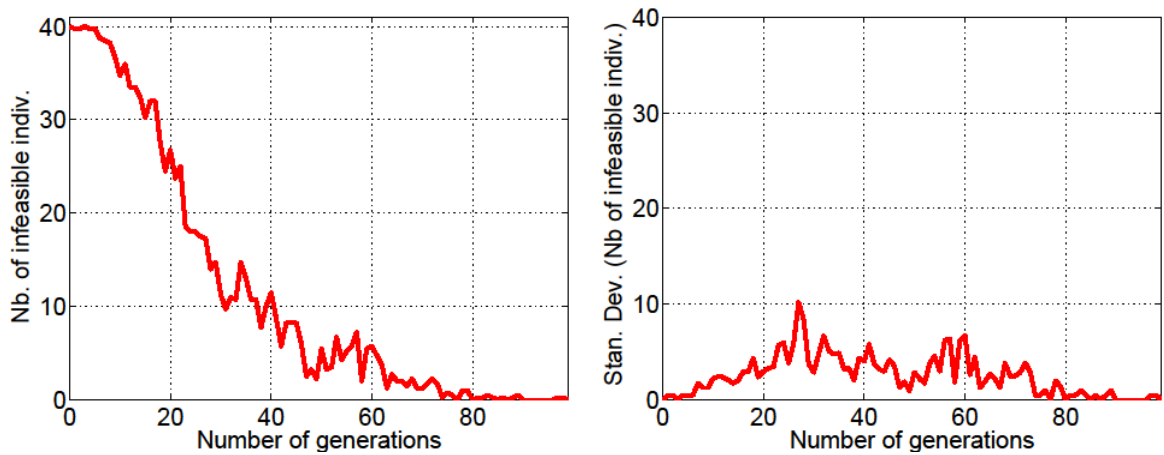

**Fig. 3**: The mean number of unfeasible individuals resampled per generation and its corresponding standard deviation using CMA-ES with an adaptive penalization with rejection technique. An individual is resampled if it is far from the feasible domain.

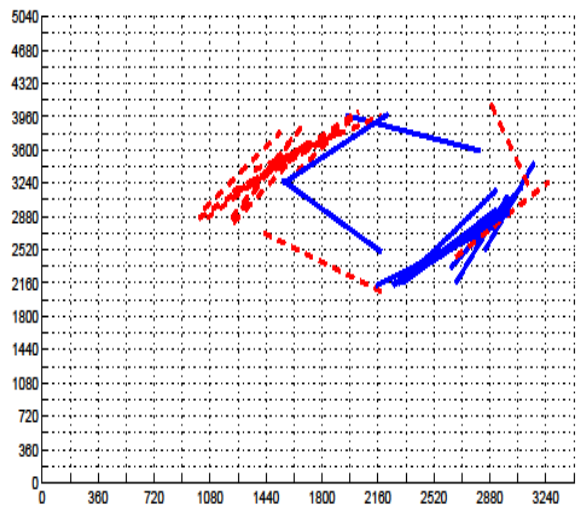

**Fig. 4**: Locations of solution wells found by 14 runs of CMA-ES projected on the top face of the reservoir. Injectors are represented by (———). Producers are represented by (- -).

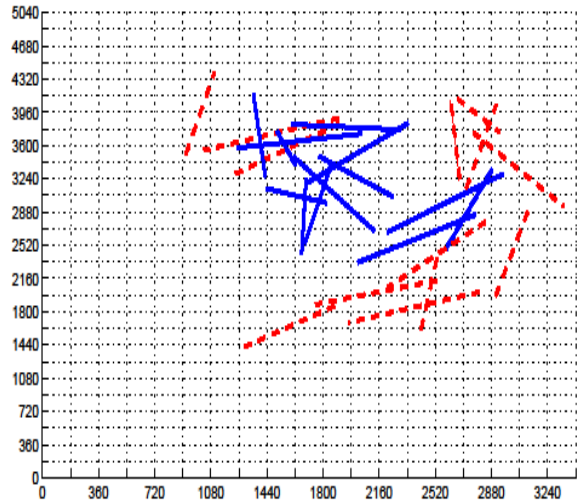

**Fig. 5**: Locations of solution wells found by 14 runs of the GA projected on the top face of the reservoir. Injectors are represented by (——). Producers are represented by (- -).

Fig. 4 shows the positions of *optimum* wells obtained from 14 runs using CMA-ES. CMA-ES succeeds in defining in 11 runs of the 14 performed the same potential region to place the producer and the injector. This region gives an NPV in between $1.99 \times 10^{10}$ \$ and $2.05 \times 10^{10}$ \$. For the other three runs, CMA-ES finds each time a different local optimum with NPV values equal to: $1.83 \times 10^{10}$ \$, $1.95 \times 10^{10}$ \$ and $2.05 \times 10^{10}$ \$. Despite the large number of local optima, CMA-ES succeeds in providing satisfactory results on 93% of the runs, if we define a satisfactory run as a run that gives an NPV greater or equal to $1.95 \times 10^{10}$ \$.

For the GA, 14 runs were performed to trace different *optimum* well configurations in Fig. 5. Well configurations are not concentrated in some regions and have an NPV mean value equal to $1.68 \times 10^{10}$ \$ with a standard deviation equal to $1.06 \times 10^{9}$. The GA leads to well configurations dispersed over a large region. The best value of NPV obtained by the GA is equal to $1.86 \times 10^{10}$ \$ and it corresponds to a well configuration close to the potential region defined by CMA-ES.

Results confirm that CMA-ES is able to find in the majority of the runs a solution in the same region. In general, CMA-ES guarantees also finding a well configuration with a satisfactory NPV value. However, the GA has a difficulty in defining these regions. The difficulty found by the GA is due to its inability to refine the obtained results. This comparison confirms the capacity of CMA-ES to better solve the well placement problem.

## 4. Meta-Models for CMA-ES

***Locally weighted regression*** To build an approximate model denoted by $\hat{f}$ of the objective function $f$, we use a locally weighted regression. During the optimization, a database, i.e., a training set is built by storing, after every evaluation on the true objective function, points together with their objective function values. Assuming that the training set contains a sufficient number $m$ of elements, for each individual, denoted now by $\mathbf{q} \in \mathrm{IR}^n$ and to be evaluated with the approximate model, we select the $k$ nearest points $(\mathbf{x}_j)_{1 \leq j \leq k}$ to $\mathbf{q}$ using the Mahalanobis distance $d$ with respect to the current covariance matrix $\mathbf{C}$ defined for a given point $\mathbf{z} \in \mathrm{IR}^n$ as $d(\mathbf{z}, \mathbf{q}) = \sqrt{(\mathbf{z}-\mathbf{q})^T \mathbf{C}^{-1}(\mathbf{z}-\mathbf{q})}$.

Locally weighted regression builds an approximate objective function using (true) evaluations $(y_j)_{1 \leq j \leq k}$ stored in the training set and corresponding to the $k$ selected nearest points to $\mathbf{q}$.

Kern et al. (2006) suggested the use of a full quadratic meta-model. Hence, using a vector $\beta \in \mathrm{IR}^{\frac{n(n+3)}{2}+1}$, we define $\hat{f}$ as follows:

```
1   approximate $\hat{f}(x_k)$, $k = 1,...,\lambda$
2   determine the $\mu$ best individuals set: $set_0$
3   determine the best individual: $elt_0$
4   evaluate f for the best individual, add to the training set
5   for $n_{ic} := 1$ to $(\lambda-1)$ do
6       approximate $\hat{f}(x_k)$, $k = 1,...,\lambda$
7       determine the $\mu$ best individuals set: $set_{n_{ic}}$
8       determine the best individual: $elt_{n_{ic}}$
9       if ( $(n_{ic}+1) < \lambda/4$ )
10          if ( ( $set_{n_{ic}} \neq set_{n_{ic-1}}$ ) or ( $elt_{n_{ic}} \neq elt_{n_{ic-1}}$ ) ) then
11              evaluate f for the best unevaluated individuals, add to the training set
12          else
13              break
14          fi
15      elseif ( $elt_{n_{ic}} \neq elt_{n_{ic-1}}$ ) then
16          evaluate f for the best unevaluated individuals, add to the training set
17      else
18          break
19      fi
20  od
```

**Fig. 6**: The approximate ranking procedure performed, for every generation, once the training set contains a sufficient number of evaluations to build the meta-model.

$$\hat{f}(\mathbf{z},\boldsymbol{\beta}) = \beta^T (z_1^2,...,z_n^2, z_1 z_2,...,z_{n-1} z_n, z_1,...,z_n,1) \quad \forall \mathbf{z} = (z_1,...,z_n) \in IR^n \; , \qquad (13)$$

The full quadratic meta-model is built based on minimizing the following criteria w.r.t. the vector of parameters $\boldsymbol{\beta}$ of the meta-model at $\mathbf{q}$:

$$A(\mathbf{q}) = \sum_{j=1}^{k} [(\hat{f}(\mathbf{x}_j,\boldsymbol{\beta}) - y_j)^2 K(\frac{d(\mathbf{x}_j,\mathbf{q})}{h})] \; , \qquad (14)$$

The kernel weighting function $K(.)$ is defined by $K(\zeta) = (1 - \zeta^2)^2$, and $h$ is the bandwidth defined by the distance of the $k^{th}$ nearest neighbor data point to $\mathbf{q}$.

In order to build the full quadratic meta-model, $k$ must be greater or equal to $\frac{n(n+3)}{2} + 1$.

***CMA-ES with local meta-models*** To incorporate the approximate model, built using the locally weighted regression, we use the approximate ranking procedure. This procedure decides whether the quality of the model is good enough in order to continue exploiting this model or new simulations should be performed. The resulting method is called the local-meta-model CMA-ES (lmm-CMA) (Kern et al., 2006). Bouzarkouna et al. (2010) proposes a new variant (nlmm-CMA) improving over lmm-CMA on most benchmark problems.

Fig. 6 gives the implementation of this procedure (denoted nlmm-CMA$_2$ in Bouzarkouna et al. (2010)). For every generation, initially, the best individual based on the meta-model is evaluated using the true objective function and then added to the training set. Another individual is evaluated until satisfying the meta-model acceptance criterion:
-   *the best individual* and *the ensemble of $\mu$ individuals* unchanged, if less than one fourth of the population is evaluated;
-   *the best individual* unchanged, if more than one fourth of the population is evaluated.

Hence, $1 + n_{ic}$ individuals are evaluated every generation where $n_{ic}$ represents the number of iteration cycles needed to satisfy the meta-model acceptance criterion.

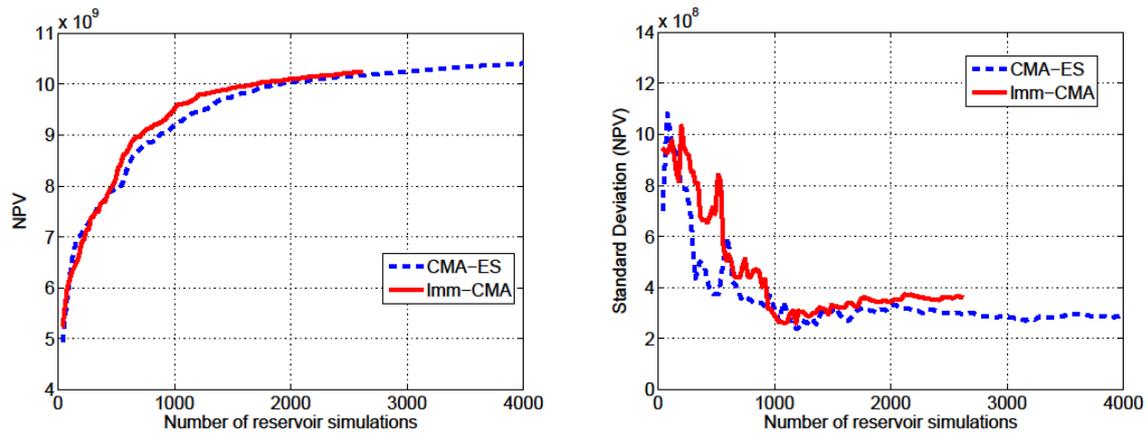

**Fig. 7**: The mean value of NPV (in US$) and its corresponding standard deviation for well placement optimization using CMA-ES with meta-models (——) and CMA-ES (- -). 10 runs are performed for each algorithm. Constraints are handled using an adaptive penalization with rejection technique.

### 5. Application of the CMA-ES with local meta-models on the PUNQ field case

In this application, we consider a placement problem of one unilateral injector and one unilateral producer. Parameters of the problem are the same as for the previous example, except for the following differences:
- a commercial reservoir simulator is used to evaluate field productions of each phase;
- the limit bottomhole pressure of the producer is fixed to 150 bar;
- the limit bottomhole pressure of the injector is fixed to 320 bar.

To define the parameters of the meta-model, we choose $k$, the number of individuals used to evaluate the meta-model, equal to 100 and meta-models are used when the training set contains at least 160 couples of points with their evaluations.

For each method, i.e. CMA-ES and CMA-ES with local meta-models (lmm-CMA), we performed 10 runs. The evolution of the NPV mean value in term of the mean number of reservoir simulations is represented in Fig. 7. Fig. 7 shows that, for the same number of reservoir simulations, combining CMA-ES with meta-models allows reaching higher NPV values compared to CMA-ES. A better representation is to show the mean number of reservoir simulations needed to reach a certain value of NPV for CMA-ES and for CMA-ES with meta-models (Fig. 8).

To reach an NPV value of $9 \times 10^9$ $, lmm-CMA, i.e. CMA-ES with meta-models requires only 659 reservoir simulations, while CMA-ES requires 880 reservoir simulations. If we consider that an NPV equal to $9 \times 10^9$ $ is satisfactory, using meta-models reduces the number of reservoir simulations by 25%. For an NPV value equal to $9.6 \times 10^9$ $, using meta-models reduces the number of reservoir simulations by 19%.

Fig. 7 and 8 highlight the contribution of meta-models in reducing the number of reservoir simulations. Results show also that, in addition to reducing the number of objective function evaluations, the method still succeed in reaching high NPV values and results are similar to those obtained by CMA-ES.

As for the previous example, the well placement optimization still succeeds in identifying in the majority of the runs the same potential region to contain *optimum* wells. In this paper, we present detailed results obtained only by one of the solution well configurations proposed by CMA-ES. The selected solution well configuration is denoted *optimized config* in the sequel. To show the contribution of the proposed approach, we compare obtained results and results obtained by 2 engineer's proposed configurations (denoted *config.1* and *config.2*). The locations and trajectories of the considered well configurations are shown in Fig. 9.

**Fig. 8**: The mean number of reservoir simulations needed to reach NPV values and its corresponding standard deviation for well placement optimization using CMA-ES with meta-models (——) and CMA-ES (- -). 10 runs are performed for each algorithm.

**Fig. 9**: Map of (H $\times \phi \times S_o$), with solution well configuration obtained using CMA-ES with meta-models (PROD-O, INJ-O) and 2 engineer's proposed well configurations (PROD-1, INJ-1 and PROD-2, INJ-2).

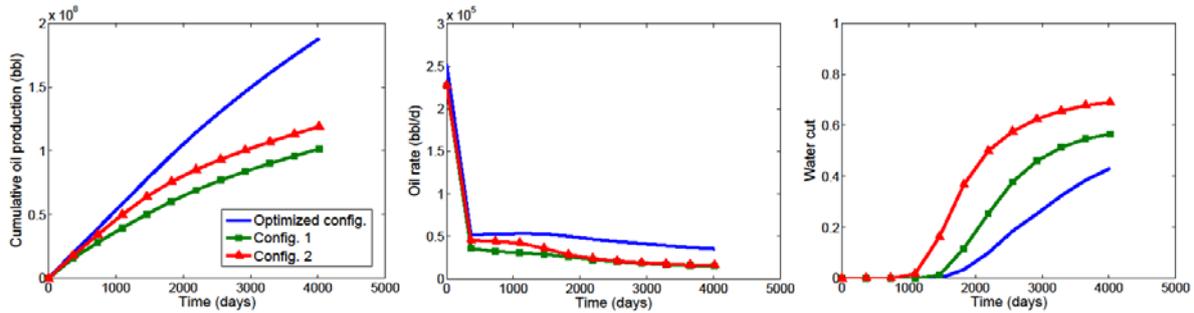

**Fig. 10**: Production curves for an optimized solution using CMA-ES with meta-models (*optimized config.*) and 2 engineer's proposed configurations (*config.1* and *config.2*).

Fig. 9 represents the distribution of $\sum_{k=1}^{n\,layers} \phi \times S_o$, where $S_o$ is the oil saturation and $\phi$ is the porosity.

PROD-*c* and INJ-*c* denotes respectively the producer and the injector corresponding to the well configuration *c*. The well configuration is either *config.1*, *config.2* or *optimized config* denoted respectively 1, 2, O. Engineer's proposed wells are horizontal wells where producers (PROD-1 = PROD-2) are placed in the top layer ($k = 1$) and injectors in the bottom layer ($k = 5$). However, producers and injectors in *optimized config* are inclined wells placed in the layer $k = 3$.

The producer proposed by engineers is placed in the region with the highest $\sum_{k=1}^{n\,layers} \phi \times S_o$ value.

Fig. 10 shows the production curves of the considered well configurations. The cumulative oil production for *optimized config*, during the 11 simulated years is equal to 189 MMbbl. While, *config.1* offers only 119 MMbbl and *config.2* offers 102 MMbbl. Hence, the proposed optimization methodology adds 59% to the best considered engineer's proposed well configuration. The *optimized config* offers also the smallest water cut (0,43 for *optimized config*, 0,57 for *config.1* and 0,69 for *config.2*).

**Conclusions**

In this work, we applied the stochastic optimization method CMA-ES to optimally place wells. A technique based on adaptive penalization with rejection was developed to handle well placement constraints with CMA-ES. The results presented show that this technique ensures that after a number of iterations, all well configurations generated by CMA-ES are either feasible or close to the feasible domain.

The optimization with CMA-ES was compared to the most popular method used in well placement optimization: the genetic algorithm, on a synthetic reservoir case. CMA-ES was shown to outperform the genetic algorithm by leading to a higher net present value (NPV). Moreover, CMA-ES is able to define potential regions containing optimal well configurations compared to the genetic algorithm that leads to a solution defining a different region for each run performed.

To tackle the computational issue related to the number of reservoir simulations performed during the optimization, the proposed methodology is coupled with local-meta-models, and then demonstrated on the reservoir case PUNQ-S3. The use of meta-models was shown to offer similar results (best NPV found and solution well configurations) as CMA-ES without meta-models. Moreover, this procedure reduces the number of simulations by 19-25% to reach a satisfactory NPV.

Comparing the obtained results with some engineer's proposed well configurations shows that the proposed optimization methodology is able to propose better well configurations in other regions that might be difficult to be determined by engineers.